\documentclass[structabstract]{aa}
\usepackage{graphicx}
\usepackage{natbib}
\begin{document}
   \title{Searches for HCl and HF in comets 103P/Hartley~2 and C/2009~P1~(Garradd) with the {\it Herschel} space observatory\thanks{{\it Herschel} is an ESA space observatory with science instruments
  provided by European-led Principal Investigator consortia and with important contribution from NASA.}
  }

\author{D.~Bockel\'ee-Morvan\inst{1} \and N.~Biver\inst{1} \and J.~Crovisier\inst{1} \and D.C.~Lis\inst{2}
\and P.~Hartogh\inst{3}  \and R.~Moreno\inst{1} \and M.~de
Val-Borro\inst{4} \and G.A.~Blake\inst{2} \and
S.~Szutowicz\inst{5} \and J.~Boissier\inst{6} \and
J.~Cernicharo\inst{7} \and S.B.~Charnley\inst{8} \and
M.~Combi\inst{9} \and M.A.~Cordiner\inst{8}  \and
T.~de~Graauw\inst{10} \and P. Encrenaz\inst{11} \and
C.~Jarchow\inst{3} \and M.~Kidger\inst{12} \and
M.~K\"{u}ppers\inst{12}   \and S.N.~Milam\inst{8} \and
 H.S.P.~M\"uller\inst{13} \and T.G.~Phillips\inst{2} \and
M.~Rengel\inst{3}}


\institute{LESIA, Observatoire de Paris, CNRS, UPMC, Universit\'e Paris-Diderot, 5 place Jules Janssen, 92195 Meudon, France\\ 
\email{dominique.bockelee@obspm.fr}
\and California Institute of Technology, Pasadena, 301-17, Pasadena, CA~91125, USA                         
\and Max-Planck-Institut f\"ur Sonnensystemforschung, Max-Planck-Str.~2, 37191 Katlenburg-Lindau, Germany           
\and Department of Astrophysical Sciences, Princeton University,
Princeton, NJ 08544, USA
\and Space Research Centre, Polish Academy of Science,  Bartycka 18a, 00-716, Warszawa, Poland                  
\and Institut de Radioastronomie Millim\'etrique, 300 rue de la
Piscine, Domaine Universitaire, 38406, Saint Martin d'H\`eres, France                                                                              
\and Department of Astrophysics, CAB, INTA-CSIC, Crta Torrej\'on-Ajalvir km 4, E-28850 Torrej\'on de Ardoz, Madrid, Spain                                                                           
\and NASA Goddard Space Flight Center, Greenbelt, MD~20770, USA                                                    
\and Department of Atmospheric, Oceanic and Space Sciences, University of Michigan, 2455 Hayward Street, Ann Arbor, MI 48109-2143, USA                                  
\and  ALMA Observatory, Alonso de C\'ordova 3107, Vitacura, Santiago, Chile                                                                 
\and LERMA, Observatoire de Paris, CNRS, UPMC, 61 avenue de l'observatoire, F-75014, Paris, France 
\and European Space Astronomy, ESAC, P.O. Box 78, 28691 Villanueva de la Ca\~nada, Spain                                     
\and I. Physikalisches Institut, Universit\"at zu K\"oln,  Z\"ulpicher Str. 77, 50937 K\"oln, Germany
         }

   \date{Received}

 \abstract{HCl and HF are expected to be the main reservoirs of fluorine and
 chlorine over a wide range of conditions, wherever hydrogen is
 predominantly molecular. They are found to be strongly
depleted in dense molecular clouds, suggesting freeze-out onto
grains in such cold environments. We can then expect that HCl and
HF were also the major carriers of Cl and F in the gas and icy
phases of the outer solar nebula, and were incorporated into
comets.} {We aimed to measure the HCl and HF abundances in
cometary ices as they can provide insights on the halogen
chemistry in the early solar nebula. } {We searched for the
$J$(1--0) lines of HCl and HF at 626 and 1232 GHz, respectively,
using the Heterodyne Instrument for the Far-Infrared (HIFI) on
board the \it{Herschel} {\rm Space} Observatory. HCl was searched
for in comets 103P/Hartley 2 and C/2009 P1 (Garradd), whereas
observations of HF were conducted in comet C/2009 P1 (Garradd). In
addition, observations of H$_2$O and H$_2$$^{18}$O lines were
performed in C/2009 P1 (Garradd) to measure the H$_2$O production
rate at the time of the HCl and HF observations. Three lines of
CH$_3$OH were serendipitously observed in the HCl receiver
setting.} {HCl is not detected, whereas a marginal (3.6-$\sigma$)
detection of HF is obtained. The upper limits for the HCl
abundance relative to water are 0.011 \% and 0.022 \%, for comet
103P/Hartley 2 and C/2009 P1 (Garradd), respectively, showing that
HCl is depleted with respect to the solar Cl/O abundance by a
factor more than 6$^{+6}_{-3}$ in 103P/Hartley 2, where the error
 is related to the uncertainty in the chlorine solar abundance.
The marginal HF detection obtained in C/2009 P1 (Garradd)
corresponds to an HF abundance relative to water of (1.8$\pm$0.5)
$\times$ 10$^{-4}$, which is approximately consistent with a solar
photospheric F/O abundance. The inferred water production rate in
comet C/2009 P1 (Garradd) is (1.1$\pm$0.3) $\times$ 10$^{29}$
s$^{-1}$ and (0.75$\pm$0.05) $\times$ 10$^{29}$ s$^{-1}$ on 17 and
23 February 2012, respectively. CH$_3$OH abundances relative to
water are 2.7$\pm$0.3 \% and $3.4\pm0.6$\%, for comets
103P/Hartley 2 and C/2009 P1 (Garradd), respectively.} {The
observed depletion of HCl suggests that HCl was not the main
reservoir of chlorine in the regions of the solar nebula where
these comets formed. HF was possibly the main fluorine compound in
the gas phase of the outer solar nebula. However, this needs to be
confirmed by future measurements.}



\authorrunning{Bockel\'ee-Morvan et al.}
\titlerunning{HCl and HF in comets}
   \maketitle
%

\section{Introduction}
Chlorine and fluorine are two of the few atoms that can react
exothermically (Cl$^+$, F) or with a small energy barrier (Cl)
with H$_2$, the dominant component of interstellar clouds and
protoplanetary disks, producing hydrogen chloride (HCl) and
hydrogen fluoride (HF), respectively, directly (Cl, F) or
eventually (Cl$^+$). Hence, based on theoretical models of
interstellar chemistry, HCl and HF are expected to be the main
reservoirs of chlorine and fluorine over a wide range of
conditions, wherever hydrogen is predominantly molecular
\citep[e.g.,][]{Neufeld2009}.

HF and HCl have both been detected in diffuse and dense molecular
clouds. Relative abundances measured in diffuse clouds show that,
indeed, fluorine is essentially locked in HF
\citep[e.g.,][]{Neufeld2010,Monje2011}, whereas HCl accounts for
less than 1\% of the elemental chlorine due to a rich chemistry
initiated by ionizing radiation \citep[e.g.,][]{Monje2013}. In
dense molecular clouds, HF and HCl are found to be strongly
depleted, which is thought to be the result of nearly all HF and
HCl molecules freezing out onto grains in such cold environments
\citep{Peng2010,Emprechtinger2012}. Both HCl and HF have
relatively high freezing points  (sublimation temperatures of 51 K
and 70 K, respectively), so they establish strong bonds onto
grains. We can then expect that HCl and HF were the major carriers
of Cl and F in the gas and icy phases of the solar nebula, and
were subsequently incorporated into comets.

The lowest rotational transitions of HCl and HF lie in the
submillimetre domain and either cannot be (HF), or can only be
with difficulty (HCl) \citep[e.g.,][]{Schilke1995}, observed from
the ground because of severe atmospheric absorption. Hence, most
detections in astrophysical sources were obtained after 2009 from
observations using the {\it Herschel} Space Observatory
\citep{Pilbratt10}. {\it Herschel} also offered
 a rare opportunity to search for these species in comets. The
solar abundances of Cl and F relative to atomic oxygen are,
respectively, (Cl/O)$_{\sun}$ $\sim$ 6.5 $\times$ 10$^{-4}$ and
(F/O)$_{\sun}$ $\sim$ 7.4 $\times$ 10$^{-5}$ \citep{Asplund2009},
so that sensitive searches required the apparitions of bright
comets during the lifetime of {\it Herschel}.

We present observations of the H$^{35}$Cl and H$^{37}$Cl $J$(1--0)
lines near 625--626 GHz in comets 103P/Hartley 2 and C/2009 P1
(Garradd) performed with HIFI \citep{2010HIFI}. A sensitive search
of HF $J$(1--0) at 1232 GHz conducted in C/2009 P1 (Garradd) was
also performed. This paper also presents HIFI observations of
several rotational lines of water undertaken in February 2012, at
the time of the halogen observations in comet C/2009 P1 (Garradd),
which complement those performed in October 2011
\citep{Bockelee2012}. Methanol (CH$_3$OH) lines observed with the
HCl receiver setting are also presented.

\begin{table*}[t]
\caption{\label{tab:1} Log of the observations.}
\begin{tabular}{lcccccccc  }
\hline\hline\noalign{\smallskip}
  Comet & Date (UT) & $r_h$ & $\Delta$$^a$ & Species/line & ObsId & HIFI  & Mode & Int. time  \\
        & yyyy/mm/dd.ddd      & (AU)   & (AU) &  & 13422\#    & band & & (min)  \\
  \hline\noalign{\smallskip}
  103P/Hartley 2       & 2010/10/30.626--30.662    & 1.059&  0.131  &  H$^{35}$Cl, H$^{37}$Cl $J$(1--0)$^{b}$    & 08589  & 1b & Point &52  \\
  C/2009P1 (Garradd)  &  2012/02/17.053--17.369    & 1.727 & 1.356 &   HF $J$(1--0)  & 39306/07   & 5a & Point & 453 \\
  C/2009P1 (Garradd)  &  2012/02/17.393--17.417    & 1.727 & 1.356 &   H$_2$O 1$_{11}$--0$_{00}$ & 39309 & 4b & Mapping  &  34 \\
  C/2009P1 (Garradd)  &  2012/02/22.745--23.004    &1.762  & 1.308  &   H$^{35}$Cl, H$^{37}$Cl $J$(1--0)$^b$  &  40372/73 & 1b & Point & 372  \\
  C/2009P1 (Garradd)  &  2012/02/23.009--23.196    & 1.763 & 1.306 &   H$_2$O 2$_{11}$--2$_{02}$ & 40375 & 2b & Point  & 269 \\
  C/2009P1 (Garradd)  &  2012/02/23.202--23.448    & 1.764 & 1.305 &  H$_2$O, H$_2$$^{18}$O 1$_{11}$--0$_{00}$& 40377 & 4b & Point  & 353 \\

 \noalign{\smallskip}\hline
\end{tabular}

{\bf Notes.} $^{(a)}$ Distance from {\it Herschel}. $^{(b)}$ also
includes the $3_2-2_1~A^-$, $13_{-1}-12_{-1}~E$, and $13_0-12_0~E$
lines of CH$_3$OH.
\end{table*}

Comet 103P/Hartley 2 is a short-period comet belonging to the
Jupiter family, which made a close approach to Earth in October
2010, and was the object of a worldwide observational campaign in
support to the EPOXI mission \citep{Ahearn2011,Meech2011}. Comet
C/2009 P1 (Garradd) is, in contrast, a long-period comet coming
from the Oort cloud, which presented significant activity at its
perihelion in December 2011. Both comets were the subjects of deep
investigations using {\it Herschel}, including measurements of the
D/H ratio in water \citep{hart11,Bockelee2012}. The observations
of 103P/Hartley 2 presented here were executed in the framework of
the guaranteed time key programme ``Water and related chemistry in
the Solar System'' \citep{hart09}. Those of C/2009 P1 (Garradd)
correspond to an open time proposal.

The HCl lines were not detected, whereas a marginal detection of
HF was obtained. We present the observations in Sect. 2, and their
analysis in Sect. 3. A discussion follows in Sect. 4.

\section{Observations}

\begin{table*}[t]
\caption{\label{tab:2} Line areas and production rates.}
\begin{tabular}{lccccll}
\hline\hline\noalign{\smallskip}
  Date (UT)$^a$ & Species & Line & $\nu$     & Beam size & Line area$^b$  & Production rate$^c$  \\
 yyyy/mm/dd.ddd          &      &          &(GHz)     & (\arcsec)   &(mK km s$^{-1}$) & (s$^{-1}$) \\
 \hline\noalign{\smallskip}
 {\it 103P/Hartley 2 } &  & & & & &\\
 2010/10/30.644 & H$^{35}$Cl & $J$(1--0)  & \phantom{1}625.9188$^d$ & 34  & $<$19$^e$ & $<$ 1.1 $\times$ 10$^{24}$ \\
 2010/10/30.644 & H$^{37}$Cl & $J$(1--0)  & \phantom{1}624.9778$^d$ & 34  & $<$18$^e$ & $<$ 1.0 $\times$ 10$^{24}$ \\
 2010/10/30.644 & CH$_3$OH & $3_2-2_1$ $A^-$  & \phantom{1}626.6263\phantom{a} & 34  & 93$\pm$3 &  (3.2$\pm$0.1) $\times$ 10$^{26}$ \\
2010/10/30.644 & CH$_3$OH & $13_0-12_0$ $E$  & \phantom{1}625.7495\phantom{a} & 34  & 16$\pm$3 & (6.4$\pm$1.3) $\times$ 10$^{26}$ \\
2010/10/30.644 & CH$_3$OH & $13_{-1}-12_{-1}$ $E$  & \phantom{1}627.1705\phantom{a} & 34  & 17$\pm$3 &  (4.8$\pm$0.8) $\times$ 10$^{26}$ \\
 & & & && &\\
{\it C/2009 P1 (Garradd)} & & & && & \\
2012/02/17.211 & HF & $J$(1--0) &  1232.4762\phantom{a} & 17 &  66$\pm$18$^f$ & (2.0$\pm$0.5) $\times$ 10$^{25}$  \\
2012/02/22.875 & H$^{35}$Cl & $J$(1--0)  & \phantom{1}625.9188$^d$ & 34 & $<$8$^e$ & $<$ 1.29 $\times$ 10$^{25}$ \\
2012/02/22.875 & H$^{37}$Cl & $J$(1--0) & \phantom{1}624.9778$^d$& 34 & $<$9$^e$ & $<$ 1.45 $\times$ 10$^{25}$\\
2012/02/22.875 & CH$_3$OH & $3_2-2_1$ $A^-$  & \phantom{1}626.6263\phantom{a} & 34  & 39$\pm$2 &  (2.6$\pm$0.1) $\times$ 10$^{27}$ \\
2012/02/22.875 & CH$_3$OH & $13_0-12_0$ $E$  & \phantom{1}625.7495\phantom{a} & 34  & 4$\pm$2 &  (3.8$\pm$1.9) $\times$ 10$^{27}$ \\
2012/02/22.875 & CH$_3$OH & $13_{-1}-12_{-1}$ $E$  & \phantom{1}627.1705\phantom{a} & 34  & 10$\pm$2 &  (6.1$\pm$1.4) $\times$ 10$^{27}$ \\
2012/02/17.405 & H$_2$O & 1$_{11}$--0$_{00}$ & 1113.3430\phantom{a}    & 19 &6607$\pm$156   & (10.8$\pm$2.7)  $\times$ 10$^{28}$$^g$ \\
2012/02/23.327 & H$_2$O & 1$_{11}$--0$_{00}$ &  1113.3430\phantom{a}   & 19 &6313$\pm$10   & (6.85$\pm$0.02) $\times$ 10$^{28}$\\
2012/02/23.105 & H$_2$O & 2$_{11}$--2$_{02}$ &  \phantom{1}752.0331\phantom{a}   & 28 &2496$\pm$6   & (8.05$\pm$0.02) $\times$ 10$^{28}$\\
2012/02/23.327 & H$_2$$^{18}$O & 1$_{11}$--0$_{00}$ & 1101.6983\phantom{a}   & 19 &69$\pm$5  & (1.51$\pm$0.11) $\times$ 10$^{26}$ \\

\noalign{\smallskip}\hline
\end{tabular}

 {\bf Notes.} $^{(a)}$ Mean date. $^{(b)}$
Line area in main beam brightness temperature scale measured on
WBS spectra, except for HF (HRS spectrum) and H$_2$O lines (median
of WBS and HRS retrievals).  Errors correspond to 1-$\sigma$
statistical noise, and upper limits correspond to 3-$\sigma$.
Calibration errors (at most 5\%) are not included. $^{(c)}$
Assuming $T_{kin}$ = 60 K for 103P/Hartley 2, and a variable
temperature law for C/2009 P1 (Garradd) (see text). $^{(d)}$
Frequency of the main hyperfine component $F$(5/2--3/2). $^{(e)}$
Weighted average of the 3 hyperfine components. $^{(f)}$ Line area
measured in the HRS spectrum. The line area measured in the WBS
spectrum is 51$\pm$11 mK, but the retrieval is somewhat
baseline-dependent. $^{(g)}$ Determined by fitting the whole map;
the error reflects the dispersion of the retrievals at different
points in the map.


\end{table*}

The observations were performed using the HIFI instrument onboard
the {\it Herschel} Space Observatory, a 3.5-m telescope of the
European Space Agency. The observing log is presented in
Table~\ref{tab:1}.

H$^{35}$Cl and H$^{37}$Cl were searched for in comet 103P/Hartley
2 on 30 October 2010 UT, i.e., near the time of its closest
approach to Earth (20 October 2010), and perihelion (28 October
2010). The heliocentric distance at the time of the observations
was $r_h$ $\sim$ 1.06 AU, and the distance from {\it Herschel} was
$\Delta$ = 0.13 AU at the time of the observations. HF and HCl
observations in comet C/2009 P1 (Garradd) were conducted on 17 and
22--23 February 2012 UT ($r_h$ $\sim$ 1.76 AU, $\Delta$ $\sim$
1.31 AU, Table~\ref{tab:1}), respectively, i.e., about two months
after its perihelion on 23 December 2011 at perihelion distance
$q$ = 1.55 AU. Solar elongation constraints prevented us from
scheduling {\it Herschel} observations of comet Garradd at
perihelion. For comet C/2009 P1 (Garradd), the integration time
was long (6--7 h for both HCl and HF). The integration time was
less than 1 h for HCl in comet 103P/Hartley 2 (Table~\ref{tab:1}).

The H$^{35}$Cl and H$^{37}$Cl $J$(1--0) lines, at 625.9153 and
624.9751~GHz\footnote{All rest frequencies in this paper were
taken from the JPL catalog http://spec.jpl.nasa.gov/
\citep{JPL-catalog}.}, respectively \citep{HCl_1-0_rot_1971}, were
observed simultaneously, in the upper sideband of band 1b HIFI
receiver. The $I$ = 3/2 nuclear spin of $^{35}$Cl and $^{37}$Cl
splits the $J$(1--0) line into three $\Delta$$F_1$ = 0, --1, +1
hyperfine components \citep{Cazzoli2004} with statistical-weight
ratio 2:3:1, with the main component $F_1$(5/2--3/2) being at the
rest frequencies of 625.9188 GHz and 624.9778 GHz for H$^{35}$Cl
and H$^{37}$Cl, respectively. The outer two components are
separated by --6.35 and +8.22 km s$^{-1}$, respectively, from the
strongest middle component for H$^{35}$Cl, and by --5.05 km
s$^{-1}$ and +6.45 km s$^{-1}$, respectively, for H$^{37}$Cl. This
splitting has been resolved in astronomical observations
\citep[e.g.,][]{Peng2010}. The much smaller splitting caused by
the $I = 1/2$ spin of the H nucleus can be resolved in the
laboratory \citep{HCl_HFS_dipole_1970,Cazzoli2004}, but not in
astronomical observations.

The HF $J$(1--0) line at 1232.4762 GHz \citep{Nolt1987} was
observed in the upper side band of the band 5a receiver.

The 1$_{11}$--0$_{00}$ para lines of H$_2$O and H$_2$$^{18}$O at
1113.3430 \citep{H2O_rot_up_to_triade1_2012}  and 1101.6983 GHz
\citep{H2O-18_rot_1972,water_isos_FIR_1985}, respectively, as well
as the 2$_{11}$--2$_{02}$ para H$_2$O line \citep[752.0331
GHz,][]{JCP-PhD_1995}, were observed at the dates of the HCl and
HF observations of comet C/2009 P1 (Garradd) (Table~\ref{tab:1}).
Except for the observations conducted on 17 February, a long
integration time was requested for these observations to possibly
detect lines of H$_2$O$^+$. These H$_2$O$^+$ observations are not
further discussed in the present paper.

Spectra were acquired with both the Wide Band Spectrometer (WBS)
and High Resolution Spectrometer (HRS). The spectral resolution of
the WBS is 1.1 MHz. The HRS was used in the nominal-resolution
mode (250 kHz spectral resolution).

All lines were observed in the two orthogonal horizontal and
vertical polarizations, and all observations, except one (H$_2$O
mapping), were performed in the Single-Point mode
(Table~\ref{tab:1}). To cancel the background radiation, the
observations of HCl and HF were carried out in the
frequency-switching observing mode (FSW) with a frequency throw of
94.5 MHz. Dual beam-switching was used for Single-Point
observations of water. The mapping observation of the
1$_{11}$--0$_{00}$ water line at 1113 GHz undertaken on 17
February 2012 covers a field of view of 3\arcmin$\times$3\arcmin.
The Half Power Beam Width (i.e., beam size) for each setting is
given in Table~\ref{tab:2}.

Comet C/2009 P1 (Garradd) was tracked using the latest ephemeris
available from the JPL Horizons system, and the estimated pointing
error is less than 2--3\arcsec~r.m.s. On the other hand, it turned
out that the ephemeris available for 103P/Hartley 2 at the time of
uploading the comet positions to the spacecraft resulted in a
pointing error of 10\arcsec, corresponding to 30\% of the beam
size  at the frequency of the HCl lines (Table~\ref{tab:2}). This
pointing offset is taken into account when determining the
production rate from the line area (the model can compute expected
line intensities for beams not centred on the nucleus).

\begin{figure}[t]
\includegraphics[width=4cm,bb = 91 40 300 670]{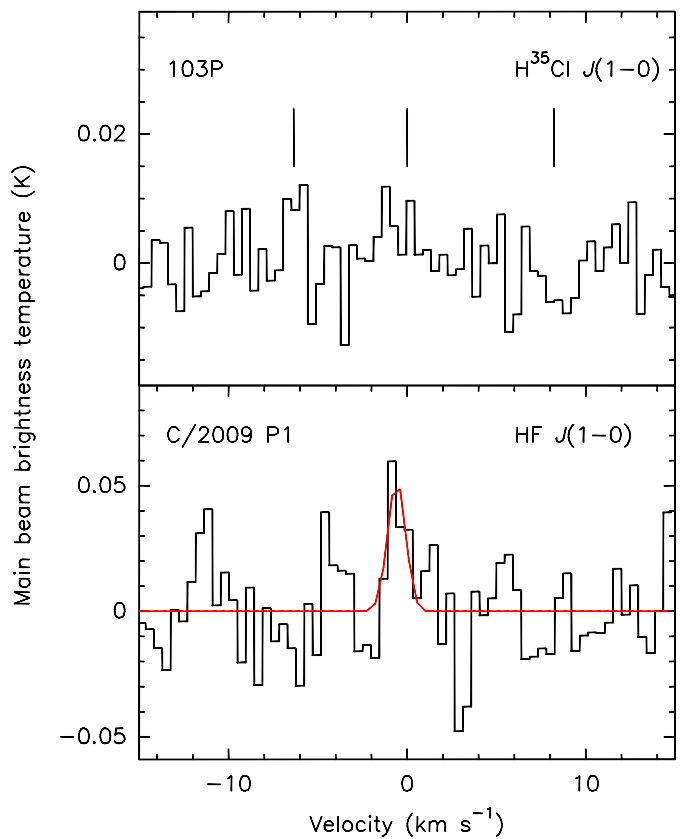}
 \caption{Spectra of the H$^{35}$Cl and  HF $J$(1--0) lines observed in comets 103P/Hartley 2 and C/2009 P1 (Garradd) on 30 October 2010
 and 17 February 2012, respectively, with the HRS. The HCl and HF spectra have been smoothed to a spectral resolution of
 0.96 MHz (0.46 km s$^{-1}$) and 1.9 MHz (0.47 km s$^{-1}$), respectively. The position of the three H$^{35}$Cl hyperfine components
 is indicated on the H$^{35}$Cl spectrum. The red curve superimposed on the HF spectrum is a Gaussian fit to the line.
 The horizontal axis is the Doppler velocity based on the rest frequency
of the transition.} \label{fig:1}
\end{figure}

\begin{figure}[t]
\begin{center}
\hspace{-2cm}
\includegraphics[width=12cm,bb = 50  50  575 660]{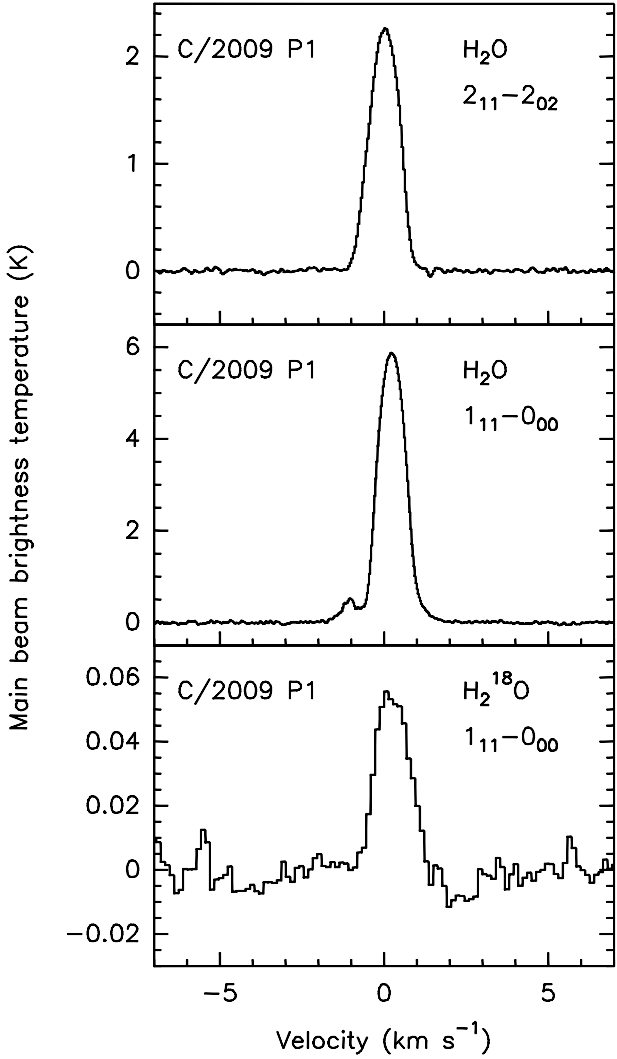}
\end{center}
 \caption{H$_2$O and H$_2$$^ {18}$O lines observed in comet C/2009 P1 (Garradd) on 22 February 2012.
 The channel spacing of the  1$_{11}$--0$_{00}$ and 2$_{11}$--2$_{02}$ H$_2$O HRS spectra is 120 kHz, corresponding to
 0.033 and 0.048 km s$^{-1}$, respectively. The WBS spectrum of the 1$_{11}$--0$_{00}$ H$_2$$^ {18}$O line has a
 spectral resolution of 1 MHz (0.27 km s$^{-1}$). The horizontal axis is the Doppler velocity based on the rest frequency
of the transition.} \label{fig:2}
\end{figure}

\begin{figure}[t]
\includegraphics[width=11cm,bb = 50  17 575 820]{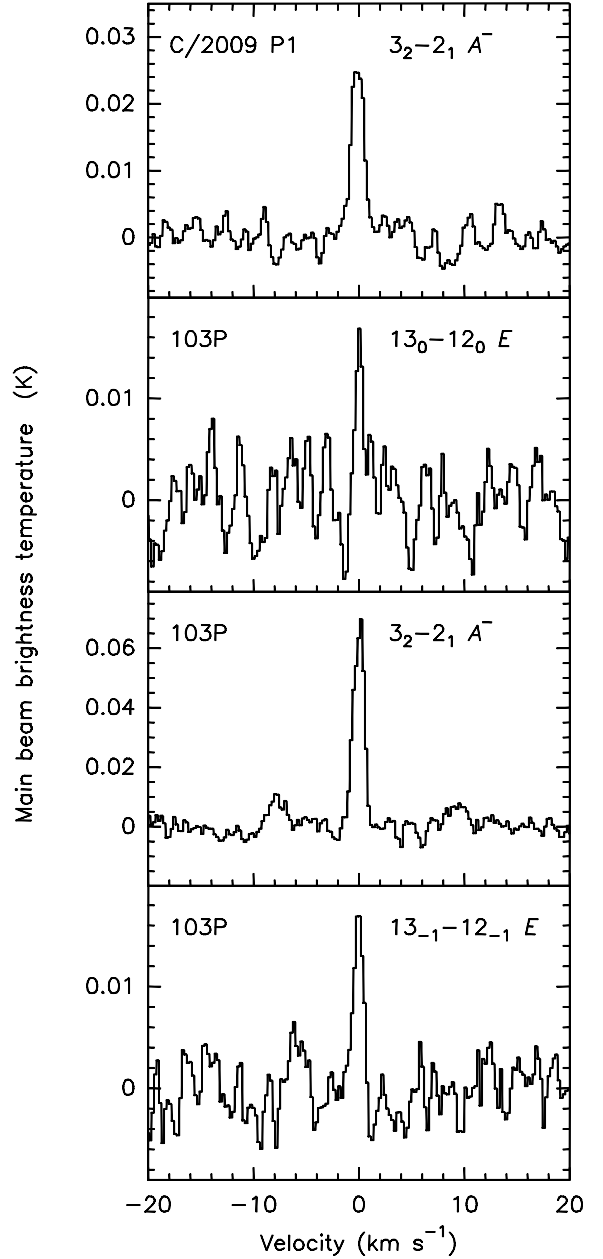}
 \caption{CH$_3$OH lines observed in WBS spectra of comets 103P/Hartley 2 and C/2009 P1 (Garradd) observed on
 30 October 2010 and 22 February 2012, respectively. The spectral resolution is 1 MHz (0.24 km s$^{-1}$). The horizontal axis is the Doppler velocity based on the rest frequency
of the transition.} \label{fig:3}
\end{figure}

\begin{figure}[t]
\begin{center}
\includegraphics[width=9.5cm,bb = 120 200 530 600]{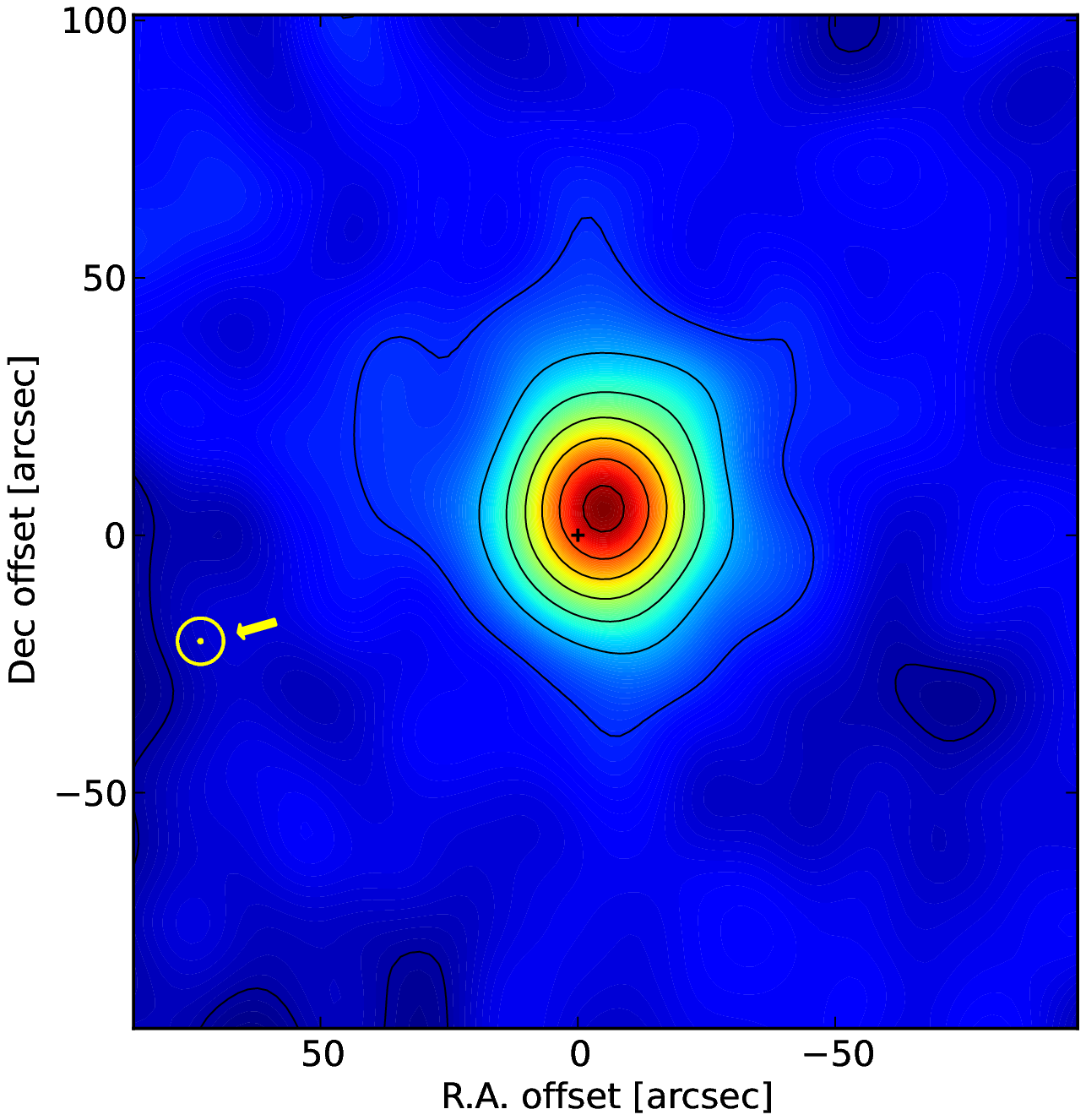}
\end{center}
 \caption{On-the-fly map of the $1_{11}$--1$_{00}$ para H$_2$O line at 1113 GHz in comet C/2009 P1 (Garradd) obtained with the WBS on 17 February
 2012.  The contour spacing is 0.6 K km s$^{-1}$ in main beam brightness temperature scale. The cross is at (0,0) coordinates. The Sun direction is indicated by the arrow. The beam size is 19\arcsec.
} \label{fig:maps}
\end{figure}

The data were first processed with the HIPE software
\citep{Ott2010}, to obtain calibrated level-2 data products, and
then exported to CLASS\footnote{http://www.iram.fr/IRAMFR/GILDAS}.
Vertical and horizontal polarizations were averaged, weighted by
the root mean square amplitude, in order to increase the signal-to
noise ratio. The main beam brightness temperature scale was
computed using a forward efficiency of 0.96, and a beam efficiency
of 0.75, 0.75, 0.74, and 0.71, for bands 1b, 2b, 4b, and 5a,
respectively. Spectra obtained after baseline removal are shown in
Figs~\ref{fig:1}--\ref{fig:3}, and the H$_2$O map is plotted in
Fig.~\ref{fig:maps}.

The HCl lines were not detected in either comet, and the
3-$\sigma$ upper limits for the line intensities, considering the
three hyperfine components with statistical-weight ratios, are
given in Table~\ref{tab:2}. A marginal detection is obtained for
HF in comet C/2009 P1 (Garradd) in both the WBS (4.6-$\sigma$ for
the line area) and HRS (3.6-$\sigma$) spectra (Table~\ref{tab:2},
Fig.~\ref{fig:1}). However, the WBS spectrum presents strong
baseline ripples due to the use of the FSW mode, and the line area
retrieved from the WBS spectrum depends on the baseline removal.
Therefore we used the line area measured on the HRS spectrum for
the determination of the HF production rate. The HF line appears
to be somewhat blueshifted in the comet rest frame, by
--0.36$\pm$0.15 km s$^{-1}$ and --0.61$\pm$0.22 km s$^{-1}$ in the
WBS and HRS spectra, respectively.

The H$_2$O and H$_2^{18}$O lines are detected with high
signal-to-noise ratios (Figs~\ref{fig:2} and \ref{fig:maps}).

Spectra acquired with the WBS cover a large (4~GHz) bandwidth.
Three methanol lines were serendipitously detected in WBS spectra
of comet 103P/Hartley 2 acquired with the HCl setting: the
$3_2-2_1$ $A^-$ line at 626.6263~GHz, and two higher energy lines
($13_0-12_0$ $E$ and $13_{-1}-12_{-1}$ $E$) at 625.7495~GHz and
627.1705~GHz, respectively (Fig.~\ref{fig:3}, Table~\ref{tab:2}).
Frequencies of the $A^-$ and $E$ lines are from
\citet{MeOH_rot_1984} and \citet{MeOH_rot_2008}, respectively. In
comet C/2009 P1 (Garradd), the CH$_3$OH line at 626.626~GHz is
well detected (Fig.~\ref{fig:3}), whereas the 625.749~GHz and
627.171~GHz lines are only marginally detected
(Table~\ref{tab:2}).

\section{Analysis}

\subsection{Excitation models}
We developed an excitation model  to convert HCl and HF line
intensities into molecular production rates, similar to models
developed by \citet{Biver1997} \citep[see also][]{Biver1999} for
other molecules. It computes the population distribution in the
ground vibrational state $v$ = 0, taking into account collisional
excitation of the rotational levels by water molecules and
electrons, infrared excitation of the $v$~=~1 bands of vibration
of HCl (at 2886 and 2883.9 cm$^{-1}$ for H$^{35}$Cl and
H$^{37}$Cl, respectively ) and HF (3961.4 cm$^{-1}$), and
spontaneous decay. The $v$(1--0) band strengths are from
\citet{Pine1985}, and correspond to excitation rates ($g$-factors)
at 1 AU from the Sun of 2.2 $\times$ 10$^{-4}$ and 6.5 $\times$
10$^{-4}$ s$^{-1}$ for HCl and HF, respectively
\citep{Crovisier2002}. The rotational energy levels of the
fundamental and excited bands are computed using the rotational
constants provided by the JPL Molecular Spectroscopy database.
Einstein-A and B coefficients of HCl and HF rotational transitions
were computed using the dipole moments published by
\citet{HCl_HFS_dipole_1970} and \citet{HF_HFS_dipole_1970},
respectively.

The models used for analyzing the water and methanol lines
consider the same excitation processes. For water lines, radiation
trapping effects are important and considered in both excitation
and radiative transfer calculations. Details on the H$_2$O and
CH$_3$OH excitation models can be found in \citet{Bockelee2012}
and \citet{Biver2000}, and references therein.

To model collisional excitation with water molecules (i.e.,
HCl--H$_2$O, HF--H$_2$O, CH$_3$OH--H$_2$O, and H$_2$O--H$_2$O
collisions) we assume a total cross-section for de-excitation of
the rotational levels of $\sigma_c$ = 5$\times$10$^{-14}$ cm$^2$
\citep[][and references therein]{Bock1987}. For comet 103P/Hartley
2, the gas kinetic temperature $T_{kin}$ is taken equal to 60~K,
corresponding to the rotational temperature of methanol lines
(Sect.~\ref{methanol}). Observations of comet C/2009 P1 (Garradd)
are analyzed using the temperature law constrained by the
observations of water (Sect.~\ref{water}). Collisions by electrons
are treated following, e.g., \citet{zakharov2007}, with the
electron density scaling factor $x_{n_e}$ taken equal to 0.2
\citep{biv07,hart10}. The local density of the molecules is
described by the Haser model \citep{Haser1957}. We use the
photodissociation rates at $r_h$ = 1 AU for the quiet Sun
published by \citet{Huebner1992} ($\beta$(HCl) = 7.2 $\times$
10$^{-6}$ s$^{-1}$ and $\beta$(HF) = 4.3 $\times$ 10$^{-7}$
s$^{-1}$). We assume an outflow velocity of 0.65 km s$^{-1}$ for
both comets, as derived from the width of the line profiles.

\subsection{Water in C/2009 P1 (Garradd)}
\label{water}

Excitation conditions of the water molecule vary in the coma, and
depend on the temperature radial profile. Therefore, the mapping
observations of the 1$_{11}$-0$_{00}$ line at 1113~GHz and the
multiple-line observations of water can be used to constrain the
excitation in the coma of comet C/2009 P1 (Garradd). Following the
method outlined by \citet{Bockelee2012}, we obtain a better fit to
the spatial evolution of the water line intensity using a
temperature which varies with the distance to the nucleus. A
variable temperature also reduces the difference between the
production rates inferred from the 1113~GHz and 752~GHz lines
observed on 23 February. In addition, based on the mean
$Q$(H$_2$O) deduced from the 1113 and 752 GHz H$_2$O lines, the
inferred $Q$(H$_2^{16}$O)/$Q$(H$_2^{18}$O) production rate ratio
is then also closer to the value determined in the same comet in
October 2011 \citep{Bockelee2012} and to the $^{16}$O/$^{18}$O =
500 Earth value. Indeed, we derive
$Q$(H$_2^{16}$O)/$Q$(H$_2^{18}$O) = $497\pm 36$ using the variable
temperature profile, whereas the value inferred using a constant
temperature is lower (e.g., $429\pm32$ for $T_{kin}$ = 50 K). This
temperature profile has a minimum at 2500--10\,000~km, with
$T_{kin}$ decreasing from 150 K to 20~K in the first 2500~km above
the surface and increasing from 20~K at 10\,000~km to 150~K at
20\,000~km.

The measured water production rate for comet C/2009 P1 (Garradd)
is then (1.1$\pm$0.3) $\times$ 10$^{29}$~s$^{-1}$ and
(0.75$\pm$0.05) $\times$ 10$^{29}$~s$^{-1}$ for 17 and 23
February, respectively, where the former value is determined by
fitting the whole map, and the latter value is deduced from the
$Q$(H$_2$$^{18}$O) value, assuming the terrestrial
$^{16}$O/$^{18}$O = 500 ratio (Table~\ref{tab:2}). These
post-perihelion $Q$(H$_2$O) measurements, taken~60 days after
perihelion, are about two times lower than the {\it Herschel}
determination obtained~80 days before perihelion
\citep{Bockelee2012}. They are are in good agreement with values
of $\sim$ 1 $\times$ 10$^{29}$~s$^{-1}$ derived for 17 and 20
February 2012 from Ly-$\alpha$ measurements using the Solar Wind
ANisotropy (SWAN) instrument onboard the Solar Heliospheric
Observatory (SOHO) \citep{Combi2013}, and with the OH production
rate of (1.12$\pm$0.08) $\times$ 10$^{29}$~s$^{-1}$ measured with
the Nan\c{c}ay radio telescope (average of 13--25 February)
(Crovisier et al., in preparation).

The water map shown in Fig.~\ref{fig:maps} shows excess emission
towards North-West, approximately in the tail direction (at PA =
286$^{\circ}$). The offset between the position of the peak and
the nucleus position obtained from the JPL Horizons ephemeris is
about 6.5\arcsec~NW~in the WBS and HRS maps. A similar offset of
the brightness peak in the approximately anti-solar direction is
also present in the pre-perihelion water maps taken with {\it
Herschel} \citep{Bockelee2012}. This excess emission in the
anti-sunward quadrant resembles that observed for comet
103P/Hartley 2 \citep{Meech2011,Knight2013} and can be explained
by water production from an icy grain halo accelerated in the tail
direction. Other indications that sublimating icy grains were
present around C/2009 P1 (Garradd) are presented by
\citet{Paganini2012}, \citet{Combi2013},  and \citet{Disanti2014}.
The asymmetric shape of the activity curve and larger production
rate before perihelion is interpreted by \citet{Combi2013} as the
result of the presence of a substantial amount of sublimating icy
grains/chunks pre-perihelion, possibly combined with seasonal
effects. The water maps obtained with {\it Herschel} suggest a
contribution from icy grains both pre and post-perihelion.

\subsection{HCl and HF}
\label{halogens}

The derived production rates (3-$\sigma$ upper limits for
H$^{35}$Cl and H$^{37}$Cl) are given in Table~\ref{tab:2}. The
terrestrial $^{35}$Cl and $^{37}$Cl abundances are of 75.77\% and
24.23\% chlorine atoms, respectively \citep{Rosman1998}. This
isotopic ratio is used to determine 3-$\sigma$ upper limits for
the HCl production rate from the $Q$(H$^{35}$Cl) and
$Q$(H$^{37}$Cl) values. We infer $Q$(HCl) $<$ 1.34 $\times$
10$^{24}$~s$^{-1}$ and $Q$(HCl) $<$ 1.60 $\times$
10$^{25}$~s$^{-1}$, for comets 103P/Hartley 2 and C/2009 P1
(Garradd), respectively.

In Table~\ref{tab:3}, production rate measurements have been
normalized to the water production rate to constrain the HCl and
HF content in cometary ices.  Numerous measurements of the water
production rate have been obtained for 103P/Hartley 2 near the
time of EPOXI closest approach on 4 November 2010, including using
{\it Herschel}
\citep{Lis2010,biv+11dps,dell+11,com+11,Crovisier2013,hart11,mum+11,Meech2011}.
 Comet 103P/Hartley 2 displayed large, rotation-induced,
gas-activity variations, so we used contemporaneous H$_2$O data
for the normalization. We assumed the value $Q$(H$_2$O) = 1.2
$\times$ 10$^{28}$ s$^{-1}$, deduced from {\it Herschel} H$_2$O
observations performed on 30.607 and 30.623 October, i.e., just
before the HCl observations \citep[][N. Biver, personal
communication]{Lis2010}. This value is consistent with the value
deduced from {\it Odin} observations of the H$_2$O 557 GHz line
obtained at the same time as the HCl observations
\citep{biv+11dps}.

The derived $Q$(HCl)/$Q$(H$_2$O) 3-$\sigma$ upper limits are 1.1
$\times$ 10$^{-4}$ and 2.2 $\times$ 10$^{-4}$ for 103P/Hartley 2
and C/2009 P1 (Garradd), respectively. Hence, a more sensitive
upper limit for the HCl content in cometary ices is derived from
the observations of 103P/Hartley 2. For HF, we derive
$Q$(HF)/$Q$(H$_2$O) = (1.8$\pm$0.5) $\times$ 10$^{-4}$ in C/2009
P1 (Garradd), under the assumption that the marginal detected line
is real. Using instead the r.m.s. of the HRS spectrum, we derive
$Q$(HF)/$Q$(H$_2$O) $<$ 1.5 $\times$ 10$^{-4}$ (3-$\sigma$).

\subsection{Methanol in 103P/Hartley 2 and C/2009 P1 (Garradd)}
\label{methanol}

The intensities of the three observed methanol lines allow us to
determine the rotational temperature of methanol. For comet C/2009
P1 (Garradd), the high-energy $13_{0}-12_{0}$ $E$ (625.749~GHz)
and $13_{-1}-12_{-1}$ $E$ (627.171~GHz) lines are only marginally
detected (Table~\ref{tab:2}). Since the upper levels of the
transitions have similar energy and Einstein-$A$ coefficient, we
used their combined intensity ($7\pm1.6$ mK km/s) to derive a
rotational temperature of methanol of $T_{rot}$ = 57$\pm$5 K. For
comet 103P/Hartley 2, the three lines are well detected and we
infer $T_{rot}$ = 60$\pm$3 K.

Methanol production rates are given in Table~\ref{tab:2}. For
103P/Hartley 2, the mean production rate, averaging values from
the three lines is $Q$(CH$_3$OH) = (3.2$\pm$0.4) $\times$
10$^{26}$ s$^{-1}$, which is consistent with values published by
\citet{Boissier2014} for 28 October 2010, keeping in mind that the
gaseous activity of comet 103P/Hartley 2 was strongly variable
\citep{Ahearn2011,Drahus2012}. For comet C/2009 P1 (Garradd), we
infer a weighted-mean value of $Q$(CH$_3$OH) = (2.6$\pm$0.4)
$\times$ 10$^{27}$ s$^{-1}$, similar to the value measured a week
earlier with the 30-m telescope of the Institut de radioastronomie
millim\'etrique (IRAM) \citep{Biver2012}. This corresponds to
abundances relative to water of 2.7$\pm$0.3~\% and $3.4\pm0.6$~\%,
for comets 103P/Hartley 2 and C/2009 P1 (Garradd), respectively,
assuming water production rates of $1.2\times10^{28}$ and
$0.75\times10^{28}$ s$^{-1}$, respectively (see Sects~\ref{water}
and \ref{halogens}).

\begin{table}[t]
\caption{\label{tab:3} HCl and HF abundances and depletions with
respect to solar cosmic abundances.}
\begin{tabular}{lcc cc}
\hline\hline\noalign{\smallskip}  HX & Comet  &
$Q$(HX)/$Q$(H$_2$O)$^a$& (X/O)$_{\sun}$$^b$ &
Depletion  \\
\noalign{\smallskip}
 & &  ($\times$ 10$^{-4}$) & ($\times$10$^{-4}$) & factor \\
 \hline\noalign{\smallskip}
HCl & 103P & $<$ 1.1$^c$\phantom{$^c$} & 6.5$^{+6.4}_{-3.2}$ & $>$ 6$^{+6}_{-3}$\\
\noalign{\smallskip}
HCl & C/2009~P1 & $<$ 2.2$^d$\phantom{$^c$} & 6.5$^{+6.4}_{-3.2}$ & $>$ 3$^{+3}_{-1}$ \\
\noalign{\smallskip}
HF & C/2009~P1  &  1.8$\pm$0.5$^{e}$ & 0.74$^{+0.74}_{-0.37}$ & 0.4$^{+0.4}_{-0.2}$\\
 \hline\noalign{\smallskip}
\end{tabular}

{\bf Notes.} $^{(a)}$ Production rate ratio. $^{(b)}$ Solar
abundances from \citet{Asplund2009}, see text. $^{(c)}$
$Q$(H$_2$O) = 1.2 $\times$ 10$^{28}$ s$^{-1}$. $^{(d)}$ $Q$(H$_2$O)
= 0.75 $\times$ 10$^{29}$ s$^{-1}$. $^{(e)}$ $Q$(H$_2$O) =
1.1 $\times$ 10$^{29}$ s$^{-1}$.
\end{table}

\section{Discussion}

The F and Cl primordial Solar System abundances, that can be
measured in the Sun, are uncertain. The solar photospheric values
have been estimated to (F/O)$_{\sun}$ = 7.4$^{+7.4}_{-3.7}$
$\times$ 10$^{-5}$  and (Cl/O)$_{\sun}$ = 6.5$^{+6.4}_{-3.2}$
$\times$ 10$^{-4}$ \citep{Asplund2009}. However, the Cl/H value
has been measured precisely in nearby HII regions, and suggests
(Cl/O)$_{\sun}$ = (4.27$\pm$ 0.9) $\times$ 10$^{-4}$
\citep{Garcia2007}. The abundance of chlorine has also been
measured in solar flares from X-ray spectra of Cl XVI lines, from
which (Cl/O) = 11.5$^{+9.3}_{-5.2}$ $\times$ 10$^{-4}$ can be
deduced. This high value could be representative of
(Cl/O)$_{\sun}$, unless some fractionation processes separating
ions and neutrals are occurring in the solar atmosphere
\citep{Sylwester2011}.

Using the solar abundances from \citet{Asplund2009}, the
$Q$(HCl)/$Q$(H$_2$O) production rate ratio measured for
103P/Hartley 2  is a factor of at least 3 to 12 lower than the
primordial Solar System (Cl/O) abundance (Table~\ref{tab:3}). A
depletion factor larger than 2--6 is obtained for comet C/2009 P1
(Garradd) (Table~\ref{tab:3}). On the other hand, the
$Q$(HF)/$Q$(H$_2$O) production rate ratio would be approximately
consistent with the cosmic (F/O), taking into account error bars
in (F/O)$_{\sun}$ (Table~\ref{tab:3}). The only other upper limit
obtained for a halogen-bearing species in comets is for NaCl in
comet C/1995 O1 (Hale-Bopp), which value is 8 $\times$ 10$^{-4}$,
i.e., a factor of 8 higher than the $Q$(HCl)/$Q$(H$_2$O) upper
limit measured in comet 103P/Hartley
\citep{Lis1997,Crovisier2004}. It is important to note that we use
the HCl and HF abundances relative to water as an indicator of the
Cl/O and F/O abundances in cometary ices. However, other
oxygen-bearing species are present in significant abundances, such
as CO$_2$, CO, and CH$_3$OH. We therefore likely overestimate the
Cl/O and F/O abundances in cometary ices by some factor when using
the abundances relative to water. This factor is difficult to
evaluate as the CO/H$_2$O and CO$_2$/H$_2$O relative abundances
inside the nuclei are not precisely known. Comet 103P/Hartley 2 is
CO--poor \citep{Weaver2011} and CO$_2$--rich \citep{Ahearn2011},
with a $Q$(CO$_2$)/$Q$(H$_2$O) production rate ratio from the
surface of 0.6, and most of the H$_2$O production coming from
grains \citep{Fougere2013}. Comet C/2009 P1 (Garradd) has a
typical CO$_2$/H$_2$O production rate ratio of 8\%
\citep{Feaga2013} and is CO--rich, with $Q$(CO)/$Q$(H$_2$O)
varying from $\sim$ 10\% to 60\% along its orbit
\citep{Biver2012,Paganini2012,McKay2012,Feaga2013}. Overall, we
may overestimate the Cl/O and F/O abundance upper limits in
cometary ices by 30--50\%.

The Cl and F elemental abundances in the rocky phase of cometary
nuclei are not known. \citet{Ishii2008} found chlorine in one
Stardust impact track and one terminal particle, but determining
the Cl elemental abundance would require averaging many particles.
Cometary dust presents analogies with primitive carbonaceous
chondrites. In the less differentiated CI meteorites, the Cl/Si
and F/Si abundances, which are measured with high precision, are
in the low range of estimated solar photospheric values
\citep{Asplund2009}. Should these abundances be representative of
the primordial Solar System abundances, then the depletion factor
of Cl in cometary ices is $>$ 3. On the other hand, the cometary
F/O, suggested from the marginal HF detection, is higher than the
cosmic value.

In the dense interstellar medium, where conditions may be
representative of the presolar cloud, HCl and HF are found to be
strongly depleted in the gas phase with respect to solar
abundances. Since chemical models predict that HF and HCl are
formed in important quantities in the gas phase, taking up to
100\% of fluorine and several  10's of percent of chlorine
\citep{Schilke1995,Neufeld2009}, it has been suggested that these
species could be locked onto grains
\citep{Schilke1995,Peng2010,Emprechtinger2012}. However, in the
L1157-B1 shock, where species from volatile and refractory grain
components are enhanced, the HCl abundance is surprisingly low,
from which one proposed interpretation is that HCl is not the main
reservoir of chlorine in the gas and icy phase of star-forming
regions \citep{Codella2012}. HCl is also detected in the warm (250
K) compact circumstellar envelope or disk of CRL 2136, with a
fractional abundance consistent with the expected HCl abundance at
this temperature, and which corresponds to approximately 20\% of
the elemental chlorine abundance \citep{Goto2013}.

For the early solar nebula, thermodynamic calculations show that
HCl gas is stable at high temperatures, NaCl becomes the dominant
species at 1100 K, and Cl is sequestered in solid phase sodalite
(Na$_8$Al$_6$Si$_6$O$_{24}$Cl$_2$) between 900 and 950 K
\citep{Fegley1980}. Chlorine will condense as solid HCl hydrates
(HCl $\cdot$ 3H$_2$O) when temperatures fall below 160 K
\citep{Zolotov2007}. Pure HCl ice will condense at $\sim$ 50 K.
Alternatively, reaction kinetics may be too slow, and Cl existed
mainly as HCl gas in the solar nebula. This scenario is advocated
by \citet{Zolotov2007}, who argue that the positive correlation
between the Cl content in chondrites and the degree of aqueous
alteration is more consistent  with a delivery of chlorine as a
component of water ice. This is also supported by the abundance of
the short-lived radionuclide $^{36}$Cl in calcium-aluminium-rich
inclusions \citep{Jacobsen2011}. On the other hand,
\citet{Sharp2013a} and \citet{Sharp2013b} exclude this scenario on
the basis of the $^{37}$Cl/$^{35}$Cl isotopic ratio measured in
chondrites. Interestingly, chlorine, as well as other halogens,
are depleted on Earth by a factor of 10 relative to solar and
chondritic abundances \citep[possible interpretations for this
deficiency are discussed in][]{Sharp2013b}.

Comets likely formed in the early Solar System farther away than
asteroids in the Main Belt, though planetary migration resulted
in a general shake-up of the small-body populations
\citep{Gomes2005,Walsh2011}. Their ices are believed to be
constituted of different phases, some molecules being direct
tracers of interstellar chemistry, while others, including complex
molecules, could have been formed in the protoplanetary disk
\citep{Hincelin2013}. The observed depletion of HCl with respect
to the solar (Cl/O)$_{\sun}$ abundance in cometary ices is
consistent with HCl not being the main reservoir of chlorine in
the regions of the solar nebula where comets formed, a result
which is consistent with the observed depletion in the L1157-B1
shock \citep{Codella2012}. The observed depletion should provide
constraints for models examining the chlorine chemistry in
protoplanetary disks and in the solar nebula.

The HF/H$_2$O value derived in C/2009 P1 (Garradd) is consistent
with the  (F/O)$_{\sun}$ solar photospheric abundance, taking into
account uncertainties in (F/O)$_{\sun}$, which would suggest that
HF is the main fluorine compound in the gas phase of the outer
solar nebula. However, it is a factor of three higher than the
(F/O) solar value of 5.4 $\times$ 10$^{-4}$ derived from the
elemental composition of primitive CI meteorites. If this value is
indeed representative of the fluorine Solar System abundance, then
our marginal detection implies a large overabundance of fluorine
in cometary ices, and therefore is probably not real. Hopefully,
better constraints on the halogen content in cometary material
will be obtained from the Rosetta mission \citep{Glassmeier2007}.




\begin{acknowledgements}
HIFI has been designed and built by a consortium of institutes and
university departments from across Europe, Canada and the United
States (NASA) under the leadership of SRON, Netherlands Institute
for Space Research, Groningen, The Netherlands, and with major
contributions from Germany, France and the US. Consortium members
are: Canada: CSA, U.Waterloo; France: CESR, LAB, LERMA, IRAM;
Germany: KOSMA, MPIfR, MPS; Ireland, NUI Maynooth; Italy: ASI,
IFSI-INAF, Osservatorio Astrofisico di Arcetri-INAF; Netherlands:
SRON, TUD; Poland: CAMK, CBK; Spain: Observatorio Astronómico
Nacional (IGN), Centro de Astrobiología (CSIC-INTA). Sweden:
Chalmers University of Technology - MC2, RSS \& GARD; Onsala Space
Observatory; Swedish National Space Board, Stockholm University -
Stockholm Observatory; Switzerland: ETH Zurich, FHNW; USA:
Caltech, JPL, NHSC.

Support for this work was provided by NASA through an award issued
by JPL/Caltech. MdVB acknowledges partial support from grants NSF
AST-1108686 and NASA NNX12AH91H. S.S. was supported by polish
MNiSW funds (181/N-HSO/2008/0).
\end{acknowledgements}

\end{document}